\documentstyle[aps,prl,preprint,epsf,floats]{revtex}
\tightenlines
\newcommand{\beq}{\begin{equation}}
\newcommand{\eeq}{\end{equation}}
\newcommand{\bqa}{\begin{eqnarray}}
\newcommand{\eqa}{\end{eqnarray}}

\def\square{\vcenter{\vbox{\hrule height.4pt
          \hbox{\vrule width.4pt height8pt
          \kern8pt\vrule width.4pt}\hrule height.4pt}}}

\def\sumint{\hbox{$\sum$}\!\!\!\!\!\!\int}

\begin{document}
\preprint{
\vbox{\halign{&##\hfil\cr
        & hep-ph/0003135 \cr
&\today\cr }}}

\title{Power-Suppressed Thermal Effects from Heavy
Particles}

\author{Eric Braaten and Yu Jia}
\address{Physics Department, Ohio State University, Columbus OH 43210, USA}

\maketitle

\begin{abstract}
{\footnotesize 
In quantum field theory, heavy particles with mass $M$ 
much greater than the temperature $T$ give not only effects 
suppressed by the Boltzmann factor $e^{-M/T}$, 
but also effects suppressed by powers of $T/M$. 
We show that power-suppressed terms in equilibrium observables
arise from effective interactions among light particles 
due to virtual heavy particles.   
We study a model introduced by Matsumoto and Yoshimura in which 
heavy bosons interact only through a term that allows pair annihilation 
into light particles.
We construct an effective Lagrangian for the light field by integrating 
out the heavy field, and use it to calculate the leading power-suppressed 
terms in the energy density. 
The thermal average of the Hamiltonian density for 
the heavy field includes  a term proportional to $T^6/M^2$, 
but we show that this term can be eliminated
by a field redefinition and therefore cannot have any physical significance.
}\end{abstract}

\newpage

\section{Introduction}

One might naively expect the effects of heavy particles with mass $M$ 
much greater than the temperature $T$ 
to be suppressed by the Boltzmann factor $e^{-M/T}$.
However, in a quantum field theory, there are additional effects 
that are suppressed only by powers of $T/M$.  
In a recent series of papers \cite{my12,my3,my4},
Matsumoto and Yoshimura have argued that there are power-suppressed terms
in the number density of heavy particles.
If there were such terms, they could greatly exceed the contribution 
from the conventional Boltzmann-suppressed terms when $T \ll M$.
This would have important implications for cosmology, because it would
imply that the relic abundance of weakly-interacting massive particles 
is much larger than the conventional predictions 
based on the Boltzmann equation.
Present bounds on the energy density of the universe would then imply 
significantly tighter constraints on the properties of 
the heavy particles that may constitute the  cold dark matter.

Matsumoto and Yoshimura  have studied the power-suppressed effects 
in a simple model with two species of scalar particles, one heavy and one light.
The only interaction of the heavy particle is one that allows 
pair-annihilation into light particles.
In the first two papers in the series~\cite{my12}, Matsumoto and Yoshimura 
used the influence functional method and a Hartree approximation
to derive a quantum kinetic equation for the momentum distribution 
of heavy particles.  Their equation includes off-shell effects
associated with the thermal width of the heavy particles.
They found that the leading power-suppressed contribution to the 
heavy-particle number density was proportional to $T^{7/2}/M^{1/2}$.
In their third paper~\cite{my3}, Matsumoto and Yoshimura studied the equilibrium
number density of heavy particles and found that the leading power-suppressed
term was actually proportional to $T^6/M^3$.  
In their fourth paper~\cite{my4}, they derived a new quantum kinetic equation 
that reproduces the equilibrium result of Ref.~\cite{my3}.

Singh and Srednicki \cite{ss} have criticized Matsumoto and Yoshimura's
conclusion that there are 
power-suppressed contributions to the number density of heavy particles.
They argued that the quantum kinetic equation of Matsumoto and Yoshimura
does not properly account for the interaction energy between the 
heavy particles and the thermal bath of light particles.
They also noted that the power-suppressed terms 
found by Matsumoto and Yoshimura are actually
contributions to the number density of virtual heavy particles.
They argued that it is the number density of on-shell heavy particles 
that is relevant to the relic abundance, and this
will have the usual Boltzmann suppression.

Srednicki recently argued that Matsumoto and Yoshimura's conclusion
is the result of an inappropriate definition 
of the number density \cite{sred}.
Srednicki considered a similar model in which the real-valued heavy field 
is replaced by a complex-valued field, 
so that the heavy bosons have a conserved charge.
He showed that there was a definition of the heavy-particle number density
that had Boltzmann suppression to all orders in perturbation theory.
Srednicki also suggested that the power-suppressed contributions to the
heavy-particle energy density
should have a simple interpretation in the effective field theory
for the light particles obtained by integrating out the heavy particles.
They are contributions to the energy density of light particles coming from  
nonrenormalizable effective interactions between the light particles.

The approach followed by Matsumoto and Yoshimura has been 
to integrate out the light field to get an quantum kinetic equation 
for the heavy particles.  The power-suppressed terms are then understood 
as arising from the thermal width acquired by the heavy particle 
when the light field is integrated out.
The heavy particle no longer has a sharp energy-momentum relation, 
but is a resonance.  The power-suppressed terms come from the tail 
of the spectral function of the resonance, where the energy and momentum
of the heavy particle are both small compared to $M$.

The philosophy of effective field theories ~\cite{eft} suggests using 
the diametrically opposite strategy. Physics involving energies and momenta 
small compared to $M$ can be understood most simply by integrating
out the heavy field.  A heavy particle whose energy and momentum  
is small compared to $M$ is off its mass-shell by an amount of order
$M$.  By the uncertainty principle, it can remain in this
highly virtual state only for a time of order $1/M$.
Light fields can propagate only over distances of order $1/M$ 
in this short time.  Thus the effects of the highly virtual heavy particle
on light fields with momenta much smaller than $M$ can be taken into account
through local interactions among the light fields.  In other words,
the light fields can be described by a local effective field theory.

The effective field theory approach was used by Kong and Ravndal \cite{kr}
to compute the leading power-suppressed terms in the energy density for
QED at temperature $T \ll m_e$, where $m_e$ is the electron mass.
They first integrated out the electron field to get a low-energy 
effective Lagrangian for photons that includes the Euler-Heisenberg term.
They then computed the energy density for this effective field theory 
at temperature $T$ and found that the leading term is proportional to 
$\alpha^2 T^8/m_e^4$.  This term can be identified as a contribution 
to the photon energy density coming from the Euler-Heisenberg
term in the effective Hamiltonian for photons. 

In this paper, we use similar effective field theory
methods to study the power-suppressed thermal 
effects in the model considered by Matsumoto and Yoshimura.
We show that the power-suppressed terms in the energy density can 
indeed be interpreted as contributions to the energy density of 
light particles from nonrenormalizable effective interactions.
We also show that the term in the energy density from which
Matsumoto and Yoshimura extracted the heavy-particle number density 
can be eliminated by a field redefinition 
and therefore can not have any physical significance.

This paper is organized as follows. 
We introduce the pair-annihilation model of Matsumoto and Yoshimura 
in section II and summarize their results on contributions  
to the energy density that are suppressed by powers of $T/M$. 
In section III, we construct an effective Lagrangian for the light field 
by integrating out the heavy field.  
We compute the leading power-suppressed terms in the energy density
by differentiating the free energy density for the effective 
theory in equilibrium at temperature $T$.
In section IV, we use a field redefinition to construct an 
effective Hamiltonian for the light field.  We show that 
the thermal average of the effective Hamiltonian density reproduces 
the leading  power-suppressed terms in the energy density.
We summarize our results in section V.

\section{ Bosonic Model with pair annihilation}

The model studied by Matsumoto and Yoshimura contains two species 
of spin-zero particles: a heavy particle of mass $M$ 
described by the field $\varphi$ and a massless particle 
described by the field $\chi$.  The Lagrangian density is
\begin{eqnarray}
{\cal L} &=&  {\cal L}_\chi
+ \frac 1{2} \partial_\mu\varphi \partial^\mu\varphi -
\frac 1{2} M^2 \varphi^2 - 
\frac 1{4}\lambda \varphi^2 \chi^2,    
\label{L-MY}
\end{eqnarray}
where ${\cal L}_\chi$ is the Lagrangian for the light field:
\begin{eqnarray}  
 {\cal L}_\chi &=&  \frac 1{2} \partial_\mu\chi \partial^\mu\chi -
\frac 1{24}\lambda_{\chi} \chi^4.   
\label{L-chi}
\end{eqnarray}
We have suppressed the counterterms needed to remove 
ultraviolet divergences. The symmetry $\varphi \rightarrow-\varphi$ 
guarantees that the heavy particle is stable at zero temperature.
The heavy particles can be created or annihilated in pairs 
via the $\varphi^2\chi^2$ interaction.  Renormalizability also requires a 
$\varphi^4$ self-interaction,  but we assume that its coefficient is
much smaller than $\lambda$, so it can be neglected. 
This is a great simplification, 
because the Lagrangian is then quadratic in the heavy field $\varphi$.
The $\chi^4$ term in (\ref{L-MY}) is necessary for thermalization of the
light field, but since we are primarily interested in the effects of the 
heavy field, we will carry out explicit calculations only to 
zeroth order in $\lambda_\chi$.

The energy density is the ensemble average of the Hamiltonion density: 
$\rho = \langle {\cal H}\rangle$.
Matsumoto and Yoshimura divide the Hamiltonian density 
into three terms: 
${\cal H} = {\cal H}_\chi + {\cal H}_\varphi + {\cal H}_{\rm int}$, where
\begin{eqnarray}
{\cal H}_\chi & = & \frac1{2} {\dot{\chi}}^2 + 
\frac1{2}(\nabla \chi)^2 + 
\frac{1}{24}\lambda_\chi \chi^4 ,  
\label{H-chi}
\\
{\cal H}_\varphi & = & \frac1{2} {\dot{\varphi}}^2 + 
\frac1{2}(\nabla \varphi)^2 + \frac1{2}M^2\varphi^2,  
\label{H-phi}  
\\
{\cal H}_{\rm int} & = & \frac1{4}\lambda\varphi^2 \chi^2. 
\label{H-int}  
\end{eqnarray} 
We have suppressed the counterterms required to renormalize 
the composite operators so that their expectation values vanish 
at zero temperature. Matsumoto and Yoshimura interpreted the 
corresponding three terms in $\rho = \rho_\chi + \rho_\varphi + \rho_{\rm int}$ 
as the energy density of the ``thermal environment'', 
the energy density of the ``system'' consisting of heavy particles, 
and the interaction energy density, 
respectively.
 
At zeroth order in $\lambda$, the energy density of the heavy particles
is that of an ideal nonrelativistic gas, 
\begin{equation} \label{free}
  \rho_\varphi  =  M (MT/2\pi)^{3/2} e^{-M/T}, 
\end{equation}
which exhibits the usual Boltzman suppression. At second order in $\lambda$, 
there are terms that are suppressed only by powers of $T/M$. 
Matsumoto and Yoshimura calculated the leading power-suppressed 
contributions for each of the 3 terms in the energy density \cite{my3}:
\begin{eqnarray} 
 \delta \rho_\chi &= & -\frac{1}{69120} \lambda^2 \frac{T^6}{M^2},  
\label{rho-chi}  
\\
  \delta \rho_\varphi &= & \frac{1}{69120} \lambda^2 \frac{T^6}{M^2},
\label{rho-phi}
\\
  \delta \rho_{\rm int} &=& -\frac{\pi^2}{64800} \lambda^2 \frac{T^8}{M^4}.
\label{rho-int}  
\end{eqnarray}
The terms proportional to $\lambda^2 T^6$ terms cancel between 
$\delta\rho_\chi$ and $\delta\rho_\varphi$ and, so the leading 
power-suppressed term of order $\lambda^2$ in the total energy
density  is proportional to  $\lambda^2 T^8$.  
Matsumoto and Yoshimura noted this cancellation, but nevertheless
interpreted $\delta\rho_\varphi$ in (\ref{rho-phi}) as a  
contribution to the energy density of heavy particles. 
Taking the heavy particles to be nonrelativistic with energy 
equal to $M$, they identified 
$\delta\rho_\varphi/M$ as a contribution to the heavy-particle
number density proportional to $T^6/M^3$.

Singh and Srednicki \cite{ss} have argued that the separation of the 
energy density into three terms corresponding to the system, 
the environment, and interactions is reasonable if and only if
$|\rho_{\rm int}| \ll \rho_\varphi$. If this condition is not
satisfied, the coupling between the system and the environment 
is effectively strong and they cannot be clearly separated.
Note that this condition is satisfied by the power-suppressed terms
(\ref{rho-phi}) and (\ref{rho-int}) if $T \ll M$.

%Operator mixing due to the interactions makes this 
%separation of ${\cal H}$ into three terms somewhat ambiguous. 
%For example, the additive counterterms implicit in each of the 
%operators (\ref{H-chi}), (\ref{H-phi}) and (\ref{H-int}) depend on 
%both $\chi$ and $\varphi$.

\section{ Effective Lagrangian}

In this section, we construct a low-energy effective Lagrangian 
for the light field $\chi$.  This effective Lagrangian reproduces the 
zero-temperature Green functions at momentum scales much less than $M$.
The effects of the heavy fields are reproduced by nonrenormalizable 
interactions with coefficients that are suppressed by powers of $1/M^2$.

An effective Lagrangian for the light particles can be constructed by 
using functional methods to integrate out the heavy field.
This method is particularly convenient for the model of
Matsumoto and Yoshimura, 
because the Lagrangian is quadratic in the field $\varphi$.
The effective action for the light field can be defined by a 
functional integral over the heavy field:
\begin{equation}
 \exp(i S_{\rm eff}[\chi])
\equiv \int {\cal D}\varphi  \exp(i \int d^4x \; {\cal L}).
\end{equation}
This effective action shouldn't be confused with the 1PI effective action 
that generates one-particle-irreducible Green functions. 
Since ${\cal L}$ is quadratic in $\varphi$, the
functional integral can be evaluated explicitly:
\begin{equation}
 S_{\rm eff}[\chi]
 =  \int d^4 x \, {\cal L}_\chi 
 + \frac{i}{2} \ln\det \left(-\partial^2 - M^2-\frac{\lambda}{2}\chi^2 
+ i\epsilon \right).
\end{equation}
The effective action can be expanded in powers of the 
coupling constant $\lambda$:
\begin{equation}
S_{\rm eff}[\chi] 
= \int d^4 x \, {\cal L}_\chi
+\frac i{2} \ln \det (-\partial^2 - M^2+ i\epsilon)
+ \sum_{n=1}^{\infty} S_{\rm eff}^{(n)}[\chi].
\label{S-eff:sum}
\end{equation}
The term of $n$'th order in $\lambda$ is
\begin{equation}
S_{\rm eff}^{(n)}[\chi] = - {i \lambda^n \over 2^{n+1} n} {\rm tr}
	\left[  (-\partial^2-M^2+ i\epsilon)^{-1}\chi^2 \right]^n.
\label{S-n}
\end{equation}
The $\ln \det(-\partial^2 - M^2)$ term in (\ref{S-eff:sum}) can be discarded, 
because it is just a $\chi$-independent constant. 
The only contribution from
$S_{\rm eff}^{(1)}$ is the local functional $\int d^4x \, \chi^2$ 
with a divergent coefficient.  It can be absorbed into
the mass counterterm for the light field. 
The terms $S_{\rm eff}^{(n)}$ for $n \ge 2$ are nonlocal functionals of the
$\chi$ field. 

Since we are interested in light fields with characteristic 
momenta much smaller than $M$, we can use the derivative expansion 
to express $S_{\rm eff}^{(n)}$ as an infinite series of local functionals. 
The derivative expansion is illustrated in Appendix A
by expanding $S_{\rm eff}^{(2)}$ to all orders. 
The lowest derivative term from $S_{\rm eff}^{(2)}$ is a $\chi^4$ term. 
It has a divergent coefficient and can be absorbed into the counterterm 
for the $\chi^4$ term in the Lagrangian.
The remaining terms have finite coefficients suppressed
by powers of $1/M^2$. They represent nonrenormalizable interactions 
among the light fields induced by virtual heavy particles.
The effective action can now
be expressed as the integral of an effective Lagrangian: 
$S_{\rm eff}[\chi]=\int d^4x \, {\cal L}_{\rm eff}$, where
\begin{eqnarray}  
 {\cal L}_{\rm eff} &=&  {\cal L}_\chi 
 - \frac{\lambda^2}{96(4 \pi)^2 M^2} \chi^2 \partial^2 \chi^2 
 + \frac{\lambda^2}{960 (4 \pi)^2 M^4} \chi^2 (\partial^2)^2 \chi^2
- \frac{\lambda^3}{96(4 \pi)^2 M^2} \chi^6  
 +  ....  
\label{L-eff}
\end{eqnarray}
We have suppressed the counterterms required to 
remove ultraviolet divergences, and we have kept all terms 
proportional to $\lambda^m(1/M^2)^n$ with $m+n \le 4$.
If we consider an observable involving a single momentum scale $p \ll M$, 
terms in ${\cal L}_{\rm eff}$ proportional to $\lambda^m(1/M^2)^n$ 
will give effects suppressed by 
$\lambda^m (p/M)^{2n}$. The $\chi^6$ term in (\ref{L-eff})
will therefore be comparable in importance to the
$\chi^2(\partial^2)^2 \chi^2$ term if $\lambda \sim (p/M)^2$. 

A similar strategy can be used at nonzero temperature $T$
to compute power-suppressed contributions 
to equilibrium observables if $T \ll M$.
Such observables can be expressed as Euclidean functional integrals 
over the fields $\varphi$ and $\chi$ with periodic boundary conditions
in the Euclidean time direction.
To calculate the power suppressed terms, 
we would first integrate over $\varphi$, then expand in powers of $\lambda$,
then carry out the derivative expansion in powers of $1/M^2$, 
and finally integrate over $\chi$.
The first three steps reduce the problem to a calculation in
an effective theory for $\chi$, 
and the final step of integrating over $\chi$ corresponds to 
computing the thermal average in that effective theory at temperature $T$.
This strategy will reproduce all the terms 
that are suppressed by powers of $T/M$,
but it will not give any Boltzmann-suppressed terms, because
the expansion in powers of $1/M^2$ eliminates terms
with an essential singularity at $1/M = 0$.

The simplest way to compute the power-suppressed terms in 
the energy density $\rho$ 
is to calculate the corresponding terms in the free energy 
density ${\cal F}$ and then differentiate.
The power-suppressed terms in ${\cal F}$ can be obtained simply 
by computing the free energy density  at temperature $T$
for the theory defined by the effective Lagrangian (\ref{L-eff}).
The leading terms are given by vacuum diagrams 
whose only vertex is one of the power-suppressed interactions in 
(\ref{L-eff}), and they can be written
\begin{eqnarray}  
\delta {\cal F} &=& 
 {\lambda^2 \over 96(4 \pi)^2 M^2} 
 	\langle \chi^2 \partial^2 \chi^2 \rangle_{\rm free} 
- {\lambda^2 \over 960 (4 \pi)^2 M^4} 
	\langle \chi^2 (\partial^2)^2 \chi^2 \rangle_{\rm free}
\nonumber \\
&& + {\lambda^3 \over 96(4 \pi)^2 M^2} \langle \chi^6 \rangle_{\rm free} 
 +  \ldots \; ,  
\label{F-pow}
\end{eqnarray}
The angular brackets $\langle ... \rangle_{\rm free}$
denote the thermal average in the free field theory.
These thermal averages are expressed as Matsubara sum-integrals
in Appendix B.  The first term on the right hand side of (\ref{F-pow}) is zero.
The remaining two terms give 
\begin{eqnarray}
\delta {\cal F} &= & {1 \over 1024}
\left( {16 \over 225}\lambda^2 {T^4 \over M^4}
	- {25 \over 48 \pi^4} \lambda^3 {T^2 \over M^2} 
	+ \ldots \right)
	{\cal F}_{\rm free}.
\end{eqnarray}
where ${\cal F}_{\rm free} = - (\pi^2/90) T^4$ 
is the free energy of a gas of free massless bosons.
By dimensional analysis, the $\chi^2 \partial^2 \chi^2$ term in 
(\ref{F-pow}) would have given a term proportional to $\lambda^2 T^6/M^2$.
The absence of such a term is related to the cancellation of the 
$\lambda^2 T^6/M^2$ terms in the energy density 
noted by Matsumoto and Yoshimura.

Once the free energy density is known, we can derive the energy
density by differentiation: 
$\rho= -T^2\frac{\partial}{\partial T}({\cal F}/T)$. 
The leading power-suppressed terms in the energy density are 
\begin{equation}  
\delta\rho = {1 \over 1024} 
\left( {112 \over 675} \lambda^2 {T^4 \over M^4} 
	- {125 \over 144 \pi^4}\lambda^3 {T^2 \over M^2}  
	+ \ldots \right) \rho_{\rm free} , 
\label{rho-ps}
\end{equation}
where $\rho_{\rm free}= (\pi^2/30) T^4$ is the energy density 
of a free gas of massless bosons.
Srednicki calculated the $\lambda^2 (T/M)^4$ term for a similar model
with a complex-valued heavy field by direct calculation in the full theory
\cite{sred}.  It is worth noting that the $\lambda^3 (T/M)^2$ term 
is equally important if $\lambda \sim (T/M)^2$.
At very low temperature, the power-suppressed terms in (\ref{rho-ps})
dominate over the leading
Boltzman-suppressed term (\ref{free}) in the energy density 
of the heavy particles, but they represent small corrections to the 
energy density of the light particles.

\section{Effective Hamiltonian}

Srednicki \cite{sred} argued that the power suppressed terms 
in the energy density should be interpreted as contributions 
to the energy density of the light field from 
nonrenormalizable effective interactions.
In order to verify this explicitly,
we construct a low-energy effective Hamiltonian density
for the light field and compute its thermal average 
in the effective theory at temperature $T$.  

If the Lagrangian density  depends only on 
the field $\chi$ and its first derivatives, the standard Noether 
prescription for constructing the Hamiltonian density is
${\cal H}= \dot{\chi}(\partial{\cal L}/\partial{\dot \chi}) - {\cal L}$. 
This prescription can not be applied to the effective Lagrangian
(\ref{L-eff}) because, even after using integration by parts to reduce 
the number of derivatives acting on any single field, it still 
depends on the second derivatives of $\chi$.  While the Noether prescription 
for the Hamiltonian density can be generalized to higher derivative 
Lagrangians, it is rather cumbersome.

A simpler approach is to first construct a different effective Lagrangian
${\cal L}'_{\rm eff}$ that depends only on $\chi$ and its first derivatives,
and then apply the Noether prescription to it.
The effective Lagrangian ${\cal L}_{\rm eff}$ in (\ref{L-eff})
is the unique effective Lagrangian that reproduces the off-shell 
Green's functions of the full theory at low momenta, 
but there are infinitely many effective Lagrangians that reproduce 
all the physical observables at low momenta.  They include all
effective Lagrangians that can be obtained from (\ref{L-eff})
by a field redefinition.
In quantum field theory, we always have the freedom to redefine the field,
because physical quantities, such as S-matrix elements,
are invariant under field redefinitions.
In renormalizable field theories, nontrivial field redefinitions
are usually not considered, because they make the theory
superficially nonrenormalizable. 
However, effective theories are already nonrenormalizable, 
so nontrivial field redefinitions don't introduce any additional complications.
In fact they can be used to simplify the effective Lagrangian by 
removing terms that don't contribute to physical quantities.
For example, by introducing the field redefinition
$ \chi \to \chi + G(\chi)$ into the kinetic term 
$\partial_\mu \chi \partial^\mu \chi$, 
we generate additional terms that can be used to cancel any terms 
of the form $G(\chi) \partial^2\chi$. 

We use the following field redefinition to simplify 
the effective Lagrangian (\ref{L-eff}):
\begin{equation} 
\chi \; \longrightarrow \; 
\chi - \frac{\lambda^2}{72(4 \pi)^2 M^2} \chi^3
 + \frac{\lambda^2}{720 (4 \pi)^2 M^4} \partial^2 \chi^3 + \ldots \;.
%+ {\cal O}\left(\frac {\lambda^{5-n}}{M^{2n}}\right).  
\label{fr}
\end{equation}
Expanding out the derivatives and rearranging them by using integration 
by parts, our effective Lagrangian reduces to
\begin{eqnarray}  
{\cal L}^\prime_{\rm eff}
&=&  {\cal L}_\chi
+ \frac{\lambda^2}{240 (4 \pi)^2 M^4} (\partial_\mu\chi\partial^\mu \chi)^2 
- \frac{\lambda^3}{96(4 \pi)^2 M^2} \chi^6 + \ldots \; .
%+ {\cal O}\left(\frac {\lambda^{5-n}}{M^{2n}}\right).  
\label{L-eff:fr}
\end{eqnarray}
Again we have kept only those terms with a total of up to 4 powers of
$\lambda$ and $1/M^2$.  This new effective Lagrangian will not reproduce the 
low-momentum Green functions of the original theory, 
but it will reproduce all physical observables involving low momenta. 

An alternative way to derive the effective Lagrangian (\ref{L-eff:fr})
starting from the original Lagrangian (\ref{L-MY}) is by matching 
physical quantities computed in both theories.
We would begin by writing down the most general effective Lagrangian 
consistent with the symmetry $\chi \to - \chi$:
\begin{eqnarray}  
{\cal L}^\prime_{\rm eff}
&=&  {\cal L}_\chi + A \chi^2 \partial_\mu \chi \partial^\mu \chi
+ B \chi^6 + C (\partial_\mu \chi \partial^\mu \chi)^2
+ D \chi^2 \partial^2 \chi \partial^2 \chi 
+ E \chi \partial^2 \chi \partial_\mu \chi \partial^\mu \chi
+ \ldots \; .
\label{L-eff:gen}
\end{eqnarray}
We would then write down the most general field redefinition 
consistent with the symmetry:
\begin{equation} 
\chi \rightarrow \chi + a \partial^2 \chi + b \chi^3
 + c \chi \partial_\mu \chi \partial^\mu \chi 
 + d \chi^2 \partial^2 \chi + \ldots \;.
\label{fr:gen}
\end{equation} 
Inserting this field redefinition into (\ref{L-eff:gen})
and expanding it out, we would find that the coeffcients 
$b$, $c$, and $d$ could be used to set $A = D = E = 0$.
To determine the remaining coefficients, such as $B$ and $C$,
we would exploit the fact that physical quantities are invariant 
under field redefinitions.  We would compute $T$-matrix elements 
involving light particles with momenta $p \ll M$ in the full theory 
using the original Lagrangian (\ref{L-MY}) and in the effective theory 
using the effective Lagrangian (\ref{L-eff:gen}).  
By matching these $T$-matrix elements, we would deduce that 
$B$ and $C$ have the values given in (\ref{L-eff:fr}).

Having constructed the new effective Lagrangian ${\cal L}^\prime_{\rm eff}$ 
in (\ref{L-eff:fr}) that depends only upon $\chi$ and its first derivatives,
we can use the standard Noether prescription to deduce the effective 
Hamiltonian.  The effective Hamiltonian density reads 
${\cal H}_{\rm eff} = {\cal H}_\chi + {\cal H}_{\rm pow}$,
where ${\cal H}_\chi$ is given in (\ref{H-chi}) and ${\cal H}_{\rm pow}$
includes all the higher dimension operators:
\begin{eqnarray} 
{\cal H}_{\rm pow} & = & 
\frac{\lambda^2}{240 (4 \pi)^2 M^4} 
(\partial_\mu \chi \partial^\mu \chi)
\left( 3{\dot\chi}^2 + ({\nabla \chi})^2 \right) 
+ \frac{\lambda^3}{96(4 \pi)^2 M^2} \chi^6
+ ... \;.  
\label{H-eff}
\end{eqnarray} 

We can now calculate the power-suppressed terms in the energy density 
by taking the thermal average $\langle {\cal H}_{\rm eff} \rangle$ 
at temperature $T$ for the effective theory defined by (\ref{L-eff:fr}).
The energy density can be written as
$\rho = \langle {\cal H}_\chi \rangle + \langle {\cal H}_{\rm pow}\rangle$.
In $\langle {\cal H}_{\rm pow}\rangle$,
the leading power-suppressed terms $\delta \rho_{\rm pow}$
are simply the thermal averages in a free field theory  
of the operators in (\ref{H-eff}).
In $\langle {\cal H}_\chi \rangle$,
the leading power-suppressed terms $\delta \rho_\chi$
come from treating the interactions in (\ref{L-eff:fr})
as first-order perturbations.
The calculations are described in more detail in Appendix B, 
and the results are
\begin{eqnarray}
\delta \rho_\chi &=& {1 \over 1024}
\left( {16 \over 135} \lambda^2 {T^4 \over M^4} 
	- {25 \over 24 \pi^4} \lambda^3 {T^2 \over M^2} \right) 
	\rho_{\rm free}, 
\label{E-chi}
\\
\delta \rho_{\rm pow} &=& {1 \over 1024}
\left( {32 \over 675} \lambda^2 {T^4 \over M^4} 
 	+ {25 \over 144 \pi^4} \lambda^3 {T^2 \over M^2} \right) 
\rho_{\rm free},  
\label{E-pow}
\end{eqnarray}
where $\rho_{\rm free} = (\pi^2/30) T^4$.  The sum of (\ref{E-chi})
and (\ref{E-pow}) reproduces our previous result (\ref{rho-ps}).
Note that $\delta \rho_\chi$ in (\ref{E-chi})
differs from the power-suppressed terms in 
$\langle {\cal H}_\chi \rangle$
in the full theory, which include the term 
(\ref{rho-chi}) suppressed by $\lambda^2 T^2/M^2$.

If we had omitted the $\lambda^2 \chi^3$ term in the field redefinition 
(\ref{fr}), the term proportional to $\chi^2\partial^2\chi^2$ 
in the effective Lagrangian (\ref{L-eff}) would not 
have been eliminated.  There would then have been an additional term 
in the effective Hamiltonian proportional to
$\chi^2 ( {\dot\chi}^2 + (\nabla \chi)^2)$. 
Its thermal average reproduces the term $\delta \rho_\varphi$ 
in (\ref{rho-phi}) calculated by Matsumoto and Yoshimura. 
However the $\chi^2\partial^2\chi^2$ interaction term 
in the effective Lagrangian gives an additional term 
in $\langle {\cal H}_\chi \rangle$
that reproduces $\delta \rho_\chi$ in (\ref{rho-chi}). 
Since the cancelling contributions (\ref{rho-chi}) and (\ref{rho-phi})
to the energy density can be eliminated by a field redefinition, 
neither can have any physical significance.
In particular, $\delta \rho_\chi/M$ cannot be interpreted as a contribution
to the number density of heavy particles.

A field redefinition was also used by Kong and Ravandal \cite{kr}
in their calculation of the energy density for QED at $T \ll m_e$.
The effective Lagrangian obtained by integrating out the 
electron field is 
\begin{eqnarray}  
{\cal L}_{\rm eff}
&=&  - {1 \over 4} F_{\mu \nu} F^{\mu \nu}
+ {\alpha \over 60 \pi m_e^2} F_{\mu \nu} \partial^2 F^{\mu \nu}
+ {\alpha^2 \over 90 m_e^4}  \left[ (F_{\mu \nu} F^{\mu \nu})^2 
			+ {7 \over 4}(F_{\mu \nu} \tilde F^{\mu \nu})^2 \right] 
+ \ldots \; .
\label{L-eff:QED}
\end{eqnarray}
The two power-suppressed terms are called the Uehling term and the 
Euler-Heisenberg term, respectively.  The Uehling term can be 
eliminated by a field redefinition:
\begin{eqnarray}  
A_\mu \;\longrightarrow\;
A_\mu + {\alpha \over 30 \pi m_e^2} \partial^2 A_\mu 
+ \ldots \; .
\label{A:fr}
\end{eqnarray}
It therefore cannot contribute to physical quantities.
The leading power-suppressed term in the energy density
comes from the Euler-Heisenberg interactions.

\section{Conclusion}

The effective-field-theory approach provides a simple way of understanding 
the contributions to equilibrium observables that are suppressed by powers 
of $T/M$.  They arise from effective interactions among light particles 
that are induced by integrating out virtual heavy particles.  
The most economical way to compute the power-suppressed terms is to 
first construct a low-energy effective Lagrangian that describes the 
light particles at $T=0$ and then consider this effective theory in 
equilibrium at temperature $T$.

For the pair annihilation model introduced in Refs. \cite{my12,my3,my4}, 
we demonstrated explicitly that the power-suppressed terms in the 
energy density can be interpreted as contributions from the light particles.  
We used the field redefinition (\ref{fr}) to construct an effective 
Hamiltonian density ${\cal H}_{\rm eff}$ for the light field $\chi$, 
and then verified that its thermal average reproduces the 
power-suppressed terms.  The field redefinition eliminated terms 
suppressed by $\lambda^2 (T/M)^2$ from individual terms in 
$\langle {\cal H}_{\rm eff}\rangle$, which otherwise would have canceled 
only after all such terms had been added together.  The fact that all the 
$\lambda^2 (T/M)^2$ terms can be eliminated by a field redefinition 
indicates that individual terms of this form cannot have any 
physical significance.

The incorrect conclusions concerning the heavy-particle number density 
in Ref. \cite{my3} stem from the authors 
having interpreted ${\cal H}_\varphi$ in (\ref{H-phi})
literally as an operator that creates only heavy particles and 
whose thermal average therefore probes the 
number density of those particles.  
However, the operator ${\cal H}_\varphi$ also creates light particles 
through loop diagrams that involve virtual heavy particles.  
Provided the momenta of the light particles
are small compared to $M$, the loop diagram can be expressed as the 
product of a short-distance coefficient that depends on $M$ and 
a local effective operator that creates light particles.    
In the definition of the composite operator  (\ref{H-phi}),
the terms with effective operators $\chi^2$, $\dot \chi^2$, 
$(\nabla \chi)^2$ and $\chi^4$ are implicitly subtracted,
so ${\cal H}_\varphi$ creates light particles 
through higher dimension effective operators.  
Thus $\langle {\cal H}_\varphi \rangle$ receives contributions 
not only from heavy particles, but also from light particles 
created by these effective operators.
It is these latter contributions that are responsible for the 
power-suppressed terms in the energy density.
Those terms cannot be related to the number density of heavy particles
that can participate in kinetic processes,
since they involve only virtual heavy particles 
with lifetimes of order $1/M$.

The strategy of effective field theory is to integrate out heavy fields 
to get an effective theory for light fields.  
The construction of a quantum kinetic equation for heavy particles 
requires exactly the opposite strategy.  Light fields must be integrated 
out to create an effective description of the heavy particles.  
Effective field theory demonstrates convincingly that the quantum 
kinetic equations derived in Ref. [3] do not describe correctly the 
evolution of the number density of heavy particles.  
Perhaps the insights from effective field theory can be used as 
guidance for deriving the correct quantum kinetic equations.

\section*{Acknowledgments}
We thank A. Heckler and G. Steigman for bringing this problem to our attention.
We thank J.O. Andersen for useful discussions. 
This work was supported in part by the U.~S. Department of 
Energy Division of High Energy Physics (grant DE-FG02-91-ER40690).

\appendix\bigskip\renewcommand{\theequation}{\thesection.\arabic{equation}}

\section{The Derivative Expansion}
\setcounter{equation}{0}

The derivative expansion can be used to express each of the term 
$S_{\rm eff}^{(n)}$ in the effective action (\ref{S-eff:sum})
as an infinite series of local functionals. 
In this appendix, we illustrate the derivative expansion by applying it
to the term $S_{\rm eff}^{(2)}[\chi]$. 
The operator $(-\partial^2 -M^2+i\epsilon)^{-1}$ in the definition 
(\ref{S-n}) of $S_{\rm eff}^{(n)}$ corresponds to the free spin-zero 
propagator:
\begin{eqnarray}
G(x,y) & = &  
\int \frac{d^4 q}{(2\pi)^4} e^{- i q\cdot (x-y)}
\frac 1{q^2-M^2+i\epsilon}. 
\label{G}
\end{eqnarray}
The definition for $S_{\rm eff}^{(2)}$ can be written
\begin{eqnarray} \label{trans}
S_{\rm eff}^{(2)}[\chi] & = & 
- \frac{i \lambda^2}{16} \int d^4x \int d^4y \, 
	G(x,y) \chi^2(y) G(y,x) \chi^2(x). 
\end{eqnarray}
Inserting the integral expression (\ref{G}) for the propagators,
this becomes
\begin{eqnarray} 
S_{\rm eff}^{(2)}[\chi] & = & 
- \frac{i \lambda^2}{16} \int d^4x \int d^4y \, \chi^2(x) \chi^2(y) 
\int \frac {d^4p}{(2\pi)^4} e^{i p\cdot(x-y)} I(p^2), 
\label{S-2}
\end{eqnarray}
where the function $I(p^2)$ is given by
\begin{eqnarray} \label{Ip2}
I(p^2)&=& 
\int \frac {d^4q}{(2\pi)^4} \, 
\frac{1}{q^2-M^2+i\epsilon} \, \frac{1}{(q+p)^2-M^2+i\epsilon}. 
\end{eqnarray}
This integral has a logarithmic ultraviolet divergence that can be
isolated by adding and subtracting $I(0)$.
The difference between the two integrals is convergent 
and can be evaluated using the Feynman parameter method:
\begin{eqnarray} 
I(p^2) &=& 
I(0) - \frac{i}{(4\pi)^2} \int_0^1 dx
	\ln \left[ 1 - x(1-x) p^2/M^2 \right] . 
\end{eqnarray}
Assuming the integral in (\ref{S-2}) is dominated by $p^2 \ll M^2$,
we can expand the logarithm into a power series in $p^2$
and then evaluate the Feynman parameter integral to get
\begin{eqnarray} 
I(p^2) &=& 
I(0) 
+ \frac{i}{(4\pi)^2} \sum_{n=1}^{\infty}\frac{n!(n-1)!}{(2n+1)!}
\left(  \frac{p^2}{M^2} \right)^n . 
\end{eqnarray}
The function $I(p^2)$ inside the integral over $p$ in  (\ref{S-2})
can be replaced by $I(-\partial_x^2)$ outside the integral.
The integral over $p$ then reduces to $\delta^4(x-y)$,
which collapses the expression to an integral over a single coordinate $x$.
Our final result for the derivative expansion is
\begin{eqnarray}
S_{\rm eff}^{(2)}[\chi] &=& 
- \frac{i\lambda^2}{16} I(0) \int d^4x \, \chi^4  
+ \frac{\lambda^2}{16 (4\pi)^2} 
	\sum_{n=1}^{\infty} {(-1)^n n! (n-1)! \over (2n+1)! M^{2 n}}
	\int d^4x \, \chi^2 ( \partial^2 )^n \chi^2.  
\end{eqnarray}
The $\chi^4$ term has a divergent coefficient, but it can be absorbed 
into one of the counterterms in the original Lagrangian.
All the higher derivative terms have finite coefficients.

\section{Thermal sum-integrals}

The calculations of thermodynamic quantities in this paper can be reduced to 
computing thermal averages in a free field theory.
In the imaginary-time formalism, these thermal averages 
are expressed as sums over Euclidean energies 
and integrals over spacial momentum.  
We use the following notation for these sum-integrals:
\begin{equation}
\sumint_P \;=\;
T \sum_{p_4} \, \int {d^3 p\over (2 \pi)^3} \,.
\end{equation}
The Euclidean 4-momentum is $P = ({\bf p},p_4= 2 \pi n T)$, 
where $n$ is any integer.  
We also use the notation $P^2 = {\bf p}^2 + p_4^2$.
 
Many of the sum-integrals required in this paper are
thermal averages of local operators in a free field theory:
\begin{eqnarray}
\langle \chi^2 \partial^2 \chi^2 \rangle_{\rm free}
&=& 4 \sumint_P \sumint_Q {1 \over P^2}, 
\\
\langle \chi^2 (\partial^2)^2 \chi^2 \rangle_{\rm free}
&=& 4 \sumint_P  \sumint_Q 
	{(P^2)^2 + P^2 Q^2 + 2 (P \cdot Q)^2 \over P^2 Q^2},
\\
\langle \chi^6 \rangle_{\rm free} 
&=& 15 \left(\sumint_P \frac {1}{P^2}\right )^3,
\\
\langle ( \partial_\mu\chi \partial^\mu\chi )^2 \rangle_{\rm free}
&=& \sumint_P \sumint_Q 
	{ P^2 Q^2 + 2 (P \cdot Q)^2 \over P^2 Q^2},
\\
\langle ( \partial_\mu\chi \partial^\mu\chi ) 
	(\nabla \chi)^2 \rangle_{\rm free}
&=& -\sumint_P \sumint_Q 
	{ P^2 {\bf q}^2 + 2 P \cdot Q {\bf p} \cdot {\bf q}
		\over P^2 Q^2},
\\
\langle  \chi^2  
	(\nabla \chi)^2 \rangle_{\rm free}
&=& \sumint_P \sumint_Q 
	{ {\bf q}^2 \over P^2 Q^2}.
\end{eqnarray}
We also need several sum-integrals that come from computing the 
thermal average of the free-field Hamiltonian density to  
first order in the effective interactions:  
\begin{eqnarray}
\langle {\cal H}_\chi \; i \int d^4x \, 
	(\partial_\mu\chi \partial^\mu\chi )^2 \rangle_{\rm free}
&=& -2 \sumint_P \sumint_Q
	{ [P^2 Q^2 + 2 (P \cdot Q)^2] (P^2 - 2 {\bf p}^2) 
		\over (P^2)^2 Q^2},
\\
\langle {\cal H}_\chi \; i \int d^4x \, \chi^6 \rangle_{\rm free}
&=&  - 45\left(\sumint_Q \frac {1}{Q^2} \right )^2
	\sumint_P { P^2 - 2 {\bf p}^2 \over (P^2)^2},
\\
\langle {\cal H}_\chi \; i \int d^4x \, 
\chi^2 \partial_\mu\chi \partial^\mu\chi \rangle_{\rm free}
&=& \sumint_P \sumint_Q
	{ (P^2 + Q^2) (P^2 - 2 {\bf p}^2) 
		\over (P^2)^2 Q^2}.
\end{eqnarray}

We can simplify the sum integrals by averaging over angles using
$\langle p^i \rangle = 0$ and 
$\langle p^i p^j \rangle = {\bf p}^2 \delta^{ij}/3$.
This reduces the double sum-integrals to products of single sum-integrals.
The single sum-integrals that are needed are 
\begin{eqnarray}
\sumint_P P^2 &=& 0, 
\\
\sumint_P 1 &=& 0, 
\\
\sumint_P {1 \over P^2} &=& {1 \over 12} T^2, 
\\
\sumint_P {{\bf p}^2 \over P^2} &=& {\pi^2 \over 30} T^4, 
\\
\sumint_P {{\bf p}^2 \over (P^2)^2} &=& {1 \over 8} T^2, 
\\
\sumint_P {({\bf p}^2)^2 \over (P^2)^2} &=& {\pi^2\over 12} T^4.
\end{eqnarray}
We have given only the temperature-dependent terms in the sum-integrals.
The temperature-independent terms are ultraviolet divergent 
and depend on the choice of ultraviolet cutoff.
The most convenient cutoff is dimensional regularization of the 
integrals over the spatial momenta.  With this cutoff,   
the temperature-independent terms vanish.

\end{document}